\documentclass[%
 aip,
 amsmath,amssymb,
 reprint,%
]{revtex4-2}

\usepackage{graphicx}
\usepackage{dcolumn}
\usepackage{bm}

\usepackage{amsmath}
\usepackage{siunitx}
\usepackage{circuitikz}

\usepackage[utf8]{inputenc}
\usepackage[T1]{fontenc}
\usepackage{mathptmx}
\usepackage{etoolbox}

\usepackage{xcolor}

\setcitestyle{numbers,square}

\makeatletter
\def\@email#1#2{%
 \endgroup
 \patchcmd{\titleblock@produce}
  {\frontmatter@RRAPformat}
  {\frontmatter@RRAPformat{\produce@RRAP{*#1\href{mailto:#2}{#2}}}\frontmatter@RRAPformat}
  {}{}
}%
\makeatother
\begin{document}

\preprint{AIP/123-QED}

\title[]{Compact Dielectric Barrier Discharge Source for a Pulsed Supersonic Beam of Metastable Triplet H$_2$}
\author{Daniel Y. Knapp}\email{d.y.knapp@vu.nl}
\affiliation{Department of Physics and Astronomy, LaserLaB, Vrije Universiteit Amsterdam, De Boelelaan 1100, 1081 HZ Amsterdam, The Netherlands}
\author{Maximilian Beyer}\email{m.beyer@vu.nl}
\affiliation{Department of Physics and Astronomy, LaserLaB, Vrije Universiteit Amsterdam, De Boelelaan 1100, 1081 HZ Amsterdam, The Netherlands}

\date{\today}

\begin{abstract}
We present a compact dielectric barrier discharge (DBD) source for producing pulsed supersonic beams of metastable molecules, designed to mount directly onto existing gas valves without modification. Unlike plate or corona discharge sources, the DBD produces no static electric field in the expansion zone and relies on capacitively coupled microdischarges that are inherently self-limiting. 
Charge accumulation on the dielectric terminates each event within nanoseconds and locally suppresses re-ignition, so that successive microdischarges are distributed across the nozzle. Arc formation and the associated hot spots are thereby prevented, and the gas is excited uniformly over the duration of the pulse.
The source requires only a single custom ceramic nozzle and a compact printed circuit board (PCB) carrying the drive electronics, mounted directly in front of the valve. The high-voltage pulses are generated in-vacuum on the PCB from a DC supply voltage. Complete design files, circuit schematics, fabrication details, and validated simulation files are provided to allow straightforward reproduction. 

The DBD is applied to produce H$_2$ in the metastable c$^3\Pi_u^-$ state. This state can be used as an intermediate step for state-selective preparation of H$_2^+$ molecular ions by pulsed field ionization of optically prepared molecular Rydberg states, without requiring vacuum-ultraviolet radiation.  

We characterize the source by time-of-flight and Resonance-Enhanced Multiphoton Ionization (REMPI) spectroscopy, obtaining an excitation fraction on the order of $10^{-8}$ into the metastable state. The dependence of the yield on stagnation pressure and discharge voltage is consistent with collisional quenching as the dominant pressure-dependent loss.

\end{abstract}

\maketitle

\section{Introduction}

Precision spectroscopy of molecular hydrogen ions (H$_2^+$ and HD$^+$) has emerged as one of the most stringent tests of fundamental physics and a competitive method for determining the proton-to-electron mass ratio~\cite{patra2020a,alighanbari2020a}. The recent demonstration of Quantum Logic Spectroscopy (QLS) indicates the potential for further improvements beyond what has already been demonstrated~\cite{holzapfel2025a}. An experimental limitation is the difficulty of preparing H$_2^+$ in a specific rovibrational state. The two dominant production methods, electron bombardment and Resonance-Enhanced Multiphoton Ionization (REMPI), are each associated with significant drawbacks: the former yields a broad state distribution~\cite{vonbusch1972a}, while the latter, though partially state-selective, requires a deep-UV laser when exciting from the neutral ground state and is further constrained by the limited number of accessible intermediate states~\cite{schmidt2020a}.

The metastable $2$p$\pi$, $c~^3\Pi_u^-$ state of H$_2$ offers a route around both limitations. It has a lifetime exceeding $1\,~\si{\milli\second}$~\cite{lichten1960a,johnson1972a,chiu1979a,berg1994a} in the vibrational ground state and an internal energy of $94{,}941.58\,~\si{\per\centi\meter}$~\cite{ross2001a}, which places the lowest ionization threshold of H$_2$~\cite{beyer2019a} only $29{,}476\,~\si{\per\centi\meter}$ above it. High Rydberg states of H$_2$ are therefore accessible with a single near-UV photon, without requiring vacuum-ultraviolet radiation~\cite{herzberg1972a,beyer2018d} or a non-resonant two-photon transition~\cite{bjerre1985a}, which causes direct photoionization. Combined with Mass-Analyzed Threshold Ionization (MATI), this opens a practical path to the state-selective preparation of H$_2^+$~\cite{merkt1993b,knapp2026a}.

Producing metastable H$_2$ is, however, considerably more demanding than producing metastable atoms or rare-gas dimers. Excitation into the triplet manifold is spin-forbidden and proceeds only by electron exchange, which requires overlap between the projectile and a target electron. The cross-sections as a function of energy therefore peak within a factor of two of the excitation threshold, whereas the dipole-allowed singlet channels peak near $60~\si{\electronvolt}$~\cite{scarlett2021a}. Only a narrow window of the electron energy distribution contributes to triplet excitation.

Lichten first confirmed the metastability of the $c~^3\Pi_u^-$ state using electron bombardment, in which a neutral beam is crossed by an electron beam~\cite{lichten1960a}. The method allows the mean electron energy to be set near the maximum of the exchange-excitation cross section, which was the explicit design goal of the high-current, low-voltage multichannel bombarder developed by Freund to overcome the Child--Langmuir space-charge limit on the current density attainable in vacuum~\cite{freund1970a}. Bombarders of this kind have since been used by several groups~\cite{eyler1983a,eyler1986a,jozefowski1992a,jozefowski1994a}. They require, however, a thermionic cathode or a specialized emitting coating. 

Discharges are not subject to the space-charge limit, the plasma being quasi-neutral, and have been combined with supersonic expansions by seeding the discharge with electrons from a hot filament, a technique equivalent to pre-ionization that was first applied to metastable rare gases~\cite{halfmann2000a} and subsequently to H$_2$~\cite{sprecher2013a}. 
A third route avoids electron impact altogether: a fast beam of H$_2^+$ is neutralized in a vapour target, populating the $c$ state by electron exchange~\cite{debruijn1982a,debruijn1986a}. This process yields a large metastable flux of high state purity, but at keV kinetic energies, which precludes ion-trap loading.

Common to both electron-impact routes is a hot surface in the source chamber. The source presented here requires none. It is based on a dielectric barrier discharge (DBD), in which the discharge is coupled capacitively to the gas through a dielectric, so that charge accumulating on the barrier extinguishes each microdischarge within nanoseconds and no channel can develop into an arc~\cite{kogelschatz1997a}. DBDs in supersonic expansions were introduced by Luria, Lavie, and Even~\cite{luria2009a} and have since been used for rare-gas atoms, hydrogen atoms~\cite{scheidegger2022a}, fluorine radicals~\cite{stranak2021a}, and molecules including OH \cite{ploenes2016a}. Two properties are decisive here: no static potential is applied in the expansion region, which is advantageous for the subsequent preparation and field ionization of Rydberg states, and the electrodes are not exposed to the plasma, so that sputtering and the associated contamination of the beam are avoided.

The source described here is distinguished by its simplicity: it can be attached to an existing \emph{Amsterdam Piezo Valve} \cite{irimia2009a} without modification. The high-voltage pulses are generated in vacuum on a compact printed circuit board from a low-voltage DC supply, eliminating the need for high-voltage feedthroughs. The PCB and a single custom ceramic nozzle mount directly onto an unmodified commercial gas valve. We characterize the metastable yield as a function of supply voltage and stagnation pressure, confirm by REMPI spectroscopy via the $n=3$ \textit{gerade} complex that the detected metastables are H$_2$ in the $c$ state, and show that the yield is limited by collisional quenching in the dense gas at the nozzle rather than by the efficiency of the discharge. Complete design files, circuit schematics, and fabrication details are provided to allow straightforward reproduction.

\section{Metastable molecular Hydrogen}

The $2$p$\pi$, $c~^3\Pi_u$ state of H$_2$ belongs to the states dissociating to the H(1s) + H(2$\ell$) dissociation threshold. Decay to the two states that belong to the H(1s) + H(1s) dissociation threshold, $X~^1\Sigma_g^+$ and $b~^3\Sigma_u^+$, is forbidden, because of the spin-selection rule $\Delta S=0$ and the $g \leftrightarrow u$ selection rule for electric dipole (E1) transitions. The nearest state of opposite parity, $a~^3\Sigma_g^+$, lies \emph{above} the $v'=0$ level of the $c$ state, so that the only allowed radiative channel is closed from the vibrational ground state.

This protection from decay only applies to one of the two $\Lambda$-doublet components. The $c~^3\Pi_u^+$ component is coupled to the $b~^3\Sigma_u^+$ continuum by a rotational (gyroscopic) interaction \cite{lefebvre-brion2004b}, which conserves parity, and predissociates on a nanosecond timescale \cite{astashkevich2015a}. The $c~^3\Pi_u^-$ component cannot couple to $b~^3\Sigma_u^+$ in this way, and is predissociated only through much weaker spin-spin and spin-orbit interactions~\cite{berg1994a}. It is therefore the $^3\Pi_u^-$ component alone that is metastable, with a lifetime of the order of $1\,~\si{\milli\second}$~\cite{lichten1960a,johnson1972a,chiu1979a,berg1994a} in $v'=0$, limited by magnetic-dipole (M1) emission to the $b$-state continuum and by the forbidden predissociation~\cite{berg1994a}. For $v'\geq1$ the $a~^3\Sigma_g^+$ state becomes energetically accessible and the E1 channel opens, reducing the lifetime by one order of magnitude.

The fate of the triplet population produced in the discharge is governed by the following processes:

\begin{subequations}
\label{eq:triplet}
\begin{align}
\mathrm{H}_2(X ~{}^1\Sigma_g^+) + e^-
  &\;\xrightarrow{\;e^-\text{ imp.}\;}\;
  \mathrm{H}_2(c ~{}^3\Pi_u) + e^- \\[4pt]
\mathrm{H}_2(c ~{}^3\Pi_u^+)
  &\;\xrightarrow{\;\text{pred.}\;}\;
  \mathrm{H}_2(b ~{}^3\Sigma_u^+) \\[4pt]
\mathrm{H}_2(c ~{}^3\Pi_u^-;\,v'\!=\!0)
  &\;\xrightarrow{\;\text{M1}\;}\;
  \mathrm{H}_2(b ~{}^3\Sigma_u^+) + h\nu \\[4pt]
  &\;\xrightarrow{\;\text{SS/SO pred.}\;}\;
  \mathrm{H}_2(b ~{}^3\Sigma_u^+) \\[4pt]
\mathrm{H}_2(c ~{}^3\Pi_u^-;\,v'\!\geq\!1)
  &\;\xrightarrow{\;\text{E1}\;}\;
  \mathrm{H}_2(a ~{}^3\Sigma_g^+)
  \;\xrightarrow{\;\text{E1}\;}\;
  \mathrm{H}_2(b ~{}^3\Sigma_u^+) \\[4pt]
\mathrm{H}_2(b ~{}^3\Sigma_u^+)
  &\;\xrightarrow{\;\text{diss.}\;}\;
  2\,\mathrm{H}(1s)
\end{align}
\end{subequations}

Collisional quenching by ground-state H$_2$ feeds the same $a~^3\Sigma_g^+$ channel~\cite{wedding1988a}.  Because of the repulsive nature of the $b~^3\Sigma_u^+$ state, all excitations in the triplet system of H$_2$ lead to dissociation into two ground state hydrogen atoms, which distinguishes metastable H$_2$ from the species for which pulsed
discharge sources are routinely used.

Ground-state radicals and atoms are produced by dissociation of a precursor. The process is spin-allowed, and once the fragment is formed, there is nothing for it to decay into. These species are, moreover, seeded at the percent level in a heavy rare gas.  Consequently, the collision bath in which they are produced is chemically inert and the only loss channel is recombination. 

The production of metastable rare-gas atoms shares with H$_2(c~^3\Pi_u^-)$ the need for a spin-changing, exchange-mediated electron impact. However, two-body quenching of He$(2\,^3S)$ or Ne$(^3P_2)$ in its own gas is negligible. Metastable He$_2$ is the apparent counter-example, since it is a metastable \emph{molecule} produced in a discharge-driven supersonic expansion \cite{raunhardt2008a,motsch2014a}, but the resemblance is superficial. No molecule is ever excited: the spin-forbidden step takes place for a He atom, and the molecule is subsequently formed by three-body association in the dense region of the expansion. The metastable He$_2$ triplet state dissociates only into inert helium atoms. 

Metastable H$_2$ is subject to the following constraints: (i) its production requires exchange excitation, (ii) its entire triplet manifold is predissociative, and (iii) the quenching partner is the carrier gas itself. Of the electron-impact events reaching the triplet manifold, only $10$--$18~\si{\percent}$ populate the $c~^3\Pi_u$ state for mean electron energies between $3$ and $10~\si{\electronvolt}$, the remainder going predominantly into the repulsive $b~^3\Sigma_u^+$ state~\cite{janev2003a,scarlett2021a}. Half of produced c-state molecules are formed in the promptly predissociating $^3\Pi_u^+$ component, and the vertical transition from $X(v''=0)$ peaks at $v'=1$, placing only $18~\si{\percent}$ in $v'=0$.  The average cross section for destruction of $c~^3\Pi_u^-$ by ground-state H$_2$ is $7.5\times10^{-15}$~cm$^2$ \cite{tischer1985a,wedding1988a}, about three times the gas-kinetic cross section defined by the Lennard-Jones diameter $d_\text{LJ}=5.54~a_0$ \cite{gallagher1974a}.  Consequently, essentially every collision quenches.

Unlike a radical seeded in an inert carrier gas, here the target and the quenching partner are the same molecule, and their densities cannot be varied independently. Diluting H$_2$ in an inert gas at fixed stagnation pressure is equivalent to reducing the stagnation pressure.

\section{Dielectric Barrier Discharge}

Electrical discharges in supersonic expansions are most often realized as plate discharges, in which the gas breaks down between two exposed metal electrodes placed at the nozzle orifice \cite{lewandowski2004a,stranak2021a,berger2025a}.

This configuration has drawbacks that are particularly severe for the present application. A static potential applied at the nozzle might extend into the interaction region, which is incompatible with the subsequent preparation and field ionization of Rydberg states and the energy deposited in the gas heats the expansion, broadening the velocity distribution and raising the internal temperature of the beam.
This can be mitigated by pulsing the discharge for an interval shorter than the gas pulse \cite{lewandowski2004a}.
A dielectric barrier discharge achieves this intrinsically.

The barrier also removes the static field from the expansion region and prevents ion acceleration to sputtering energies. The price is efficiency: at the same backing pressure the barrier discharge yields fewer radicals than a plate discharge \cite{stranak2021a}, presumably because the gas is processed only within the microdischarge channels. In return, the barrier discharge produces internally colder products \cite{stranak2021a, ploenes2016a}, while both sources give comparably narrow velocity spreads.

A high-voltage pulse is capacitively coupled to the gas through the ceramic nozzle, which acts as the eponymous dielectric barrier and prevents arc formation by causing charge to accumulate on its surface.  The deposited charge acts in opposite senses depending on the polarity of the applied field. Within a single event, charge deposited on the dielectric cancels the field in the gas and extinguishes the microdischarge within nanoseconds, so that no channel can develop into an arc, and the same charge suppresses re-ignition at that site, forcing successive microdischarges to form elsewhere on the barrier \cite{kogelschatz1997a}. The discharge is thereby distributed over the nozzle and the gas is excited uniformly over the duration of the pulse. This is the principal advantage over a plate discharge, in which a single constricted channel can carry the entire current, producing localized heating and strongly non-uniform excitation of the beam. After the polarity of the drive waveform reverses, however, the residual surface charge adds to the applied field rather than opposing it, lowering the voltage required for the next breakdown. This memory effect sustains ignition events into the decaying tail of the drive waveform, where the instantaneous output voltage alone would be insufficient to break down the gas.

\section{Description of the Apparatus}
\begin{figure*}[htbp!]
    \includegraphics[width=0.95\linewidth]{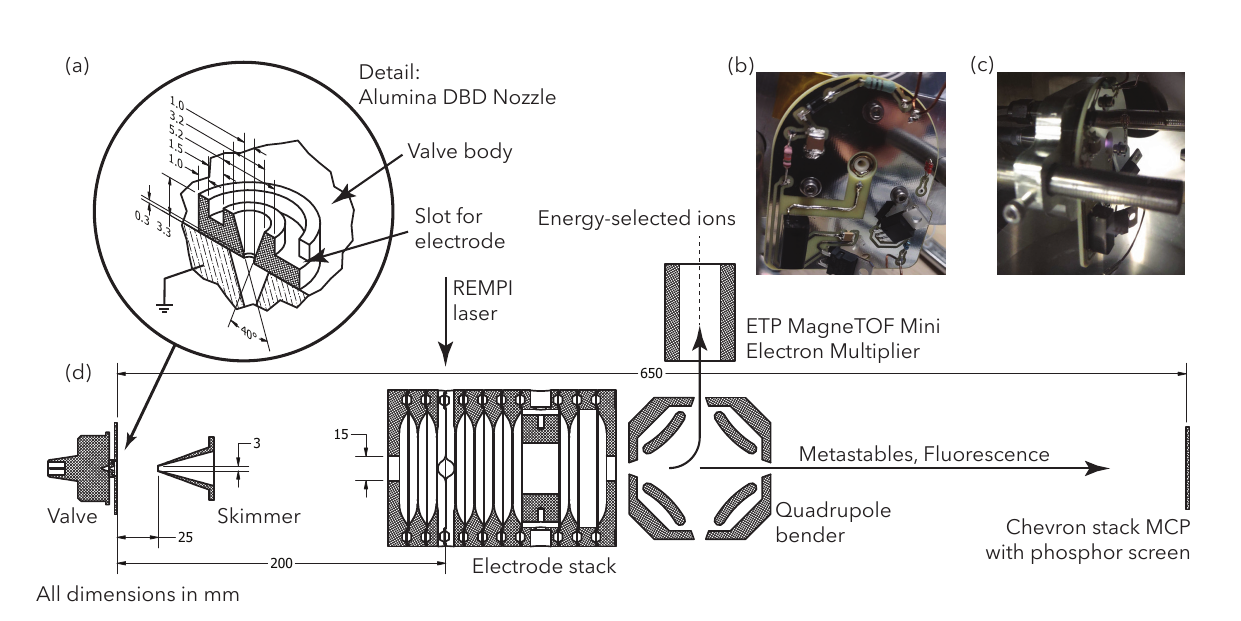}
    \caption{(a) Detail of the ceramic DBD nozzle.  (b) Image of the DBD installed in the experimental apparatus.  (c) The DBD in vacuum with visible fluorescence.  (d) Diagram of the beamline used for ToF measurements and spectroscopy.}\label{fig:discharge_cad}
\end{figure*}
The apparatus is shown in Fig.~\ref{fig:discharge_cad}.
A PCB, pictured Fig.~\ref{fig:discharge_cad}(b), is secured to the front of an Amsterdam Piezo Valve \cite{irimia2009a} (custom front plate with $800~\si{\micro\meter}$ diameter nozzle, $40^\circ$ cone opening angle) with three screws through a ceramic nozzle \cite{stcera} (detail in Fig.~\ref{fig:discharge_cad}(a)) clamped between the PCB and the front plate of the valve.
The front flange of the Amsterdam valve is secured by twelve screws, three of which were removed to reveal M$2.5$ threaded holes for clamping the PCB.
The DBD electrode is produced by passing a $1~\si{\milli\meter}$ bare copper wire through the circular slot around the nozzle, and then soldering the loose end to the PCB.
Fig.~\ref{fig:discharge_cad}(c) shows the valve during operation with H$_2$ gas, with a faint purple discharge plume visible.
Only two electrical inputs are required for operation: a power supply voltage and a trigger pulse input.
The vacuum chamber is used as a shared ground node.
The DBD drive circuit implemented on the PCB is described in Appendix A.

The vacuum outgassing rates of the FR-4 PCB substrate and the electrical components have not been rigorously tested, but they are negligible compared to the gas load of the molecular beam itself during operation, where the pressure in the source chamber is raised from its background pressure of $\num{1.0e-7}$~mbar to $\num{5.0e-6}$~mbar.
No measurable vacuum degradation attributable to the PCB or its components was observed during operation.

The valve and DBD source are placed in a separate vacuum chamber, connected to the main vacuum chamber via a $3~\si{\milli\meter}$ diameter skimmer, placed approximately $2.5~\si{\centi\meter}$ from the front of the nozzle, as shown in Fig.~\ref{fig:discharge_cad}(d).
The beam passes through a set of electrodes that produces a uniform electric field along the beam direction, with openings for laser access \cite{Ochoa2026a}.
The laser intersects the molecular beam at a distance of $20~\si{\centi\meter}$ from the valve orifice.

The beam then passes through a quadrupole bender that is used to energy-selectively deflect ions toward an electron multiplier detector placed at a right angle to the molecular beam \cite{knapp2026a}.
The non-deflected part of the beam strikes a Microchannel Plate (MCP) detector, configured with two stages in a chevron configuration.
The full distance from the front edge of the nozzle to the MCP is approximately $65~\si{\centi\meter}$.

\begin{figure}[htbp!]
    \centering
    \includegraphics[width=0.8\linewidth]{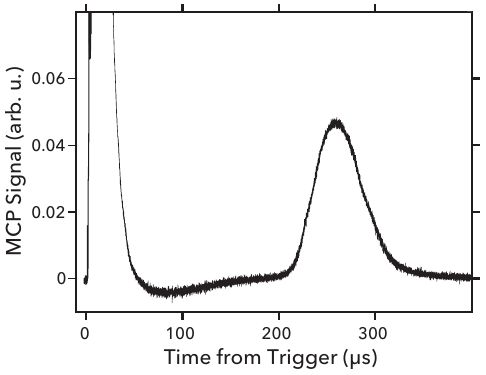}
    \caption{Full ToF measurement from discharge trigger at $t=0$ to detection of metastables on the MCP.  The vertical scale is normalized to the maximum instantaneous signal using the prompt fluorescence peak from the DBD.  Negative signal is caused by the high-pass frequency characteristic of the capacitively-coupled MCP output.}
    \label{fig:full_tof}
\end{figure}

The in-line MCP was operated with its beam-facing surface at $-1.8~\si{\kilo\volt}$, and a total potential difference of approximately $1.3~\si{\kilo\volt}$ across the two plates.
This configuration makes it sensitive to VUV photons produced by fluorescence in the initial discharge, metastable atoms and molecules, and fast ions.
Ions were blocked by the potential on the electrode stack and deflected using the quadrupole bender, so they are not detected.
To prevent fluorescence and metastables from degrading MCP surface coating, a metal plate was attached to a rotation feedthrough to block it when not in use.
A representative Time-of-Flight (ToF) measurement is shown in Fig.~\ref{fig:full_tof}.
The fluorescence signal is more than an order of magnitude stronger than the metastable signal, which limited the maximum gain at which the MCP could be operated, making the MCP generally unsuitable for the detection of H$_2^+$ ions produced by photoionization in spectroscopic measurements.
This motivated the addition of the off-axis electron multiplier, which did not show any response to fluorescence or metastables.
It could therefore be operated at a high gain, in ion-counting mode to perform a background-free measurement of the photoionization yield.

ToF spectra were recorded by momentarily lowering the electric field in the stack and turning it back on to accelerate the ions that were admitted.
However, the results were complex and could not be reliably interpreted.
The ions produced by the DBD have unknown kinetic energies, and repeated ignitions are expected to produce multiple overlapping bursts of ions.
Previous tests of DBD sources have shown that ToF spectra recorded by accelerating the ions in the transverse direction are easier to interpret \cite{scheidegger2022a, stranak2021a}.

The valve-opening duration and stagnation pressure were set to $25~\si{\micro\second}$ and $1.2$~bar respectively; lower pressures increased the metastable yield, but pressures below $1.2$~bar were avoided to reduce the risk of air contamination through gas fittings.
These parameters were chosen by optimizing the metastable yield.
The experiment was operated at a $10~\si{\hertz}$ repetition rate.
The DBD was triggered after a delay of $40~\si{\micro\second}$ relative to the valve trigger pulse.
The metastable yield was not strongly dependent on the trigger delay, and a similar signal level could be achieved over a relatively broad time window between $30~\si{\micro\second}$ and $55~\si{\micro\second}$ after the valve trigger, corresponding roughly to the temporal width of the gas pulse.

\section{Results}
\begin{figure*}[htbp!]
    \includegraphics[width=0.8\linewidth]{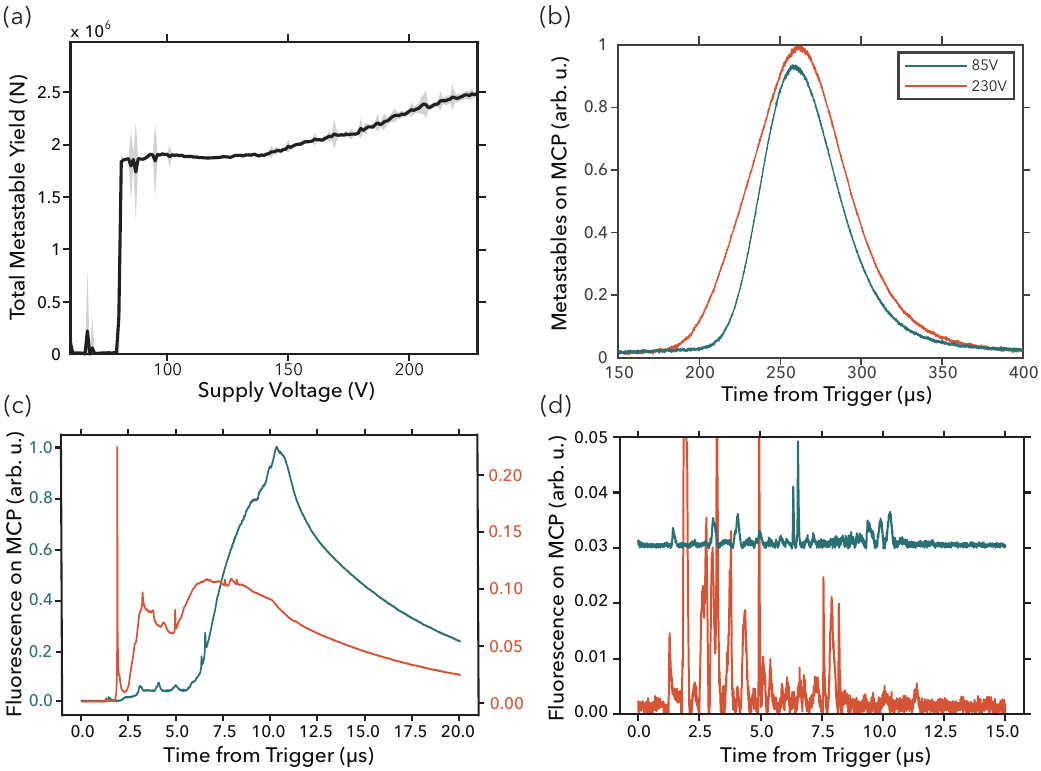}
    \caption{(a) Estimated number of metastables measured on the MCP as a function of supply voltage $V_\text{in}$, with the standard deviation of $100$ shots shown as a gray shaded region.  (b) Detail of ToF curves in the region of the metastable signal for $V_\text{in}=85~\si{\volt}$ and $230~\si{\volt}$.  (c) Fluorescence signal measured on the MCP for the same two voltages as in (b).  (d) Base-line subtracted detail of (c), showing sharp fluorescence spikes corresponding to individual DBD ignition events, with the same voltages.  The $85~\si{\volt}$ trace is shifted upward by $0.03$ for visual clarity.  Panels (c) and (d) are normalized to the same scale.}\label{fig:yield_figs}
\end{figure*}

\subsection{Beam properties}
The arrival-time profile of Fig. \ref{fig:full_tof} characterizes the expansion. The peak at $260~\si{\micro\second}$ over the $65~\si{\centi\meter}$ flight gives a beam velocity of $2500~\si{\meter\per\second}$. The profile has a width of $57~\si{\micro\second}$ FWHM (this contains in quadrature the $25~\si{\micro\second}$ pulse width at the source). This results in a speed ratio $S = 1.665\,v/\Delta v_\text{FWHM} = 8.4$, corresponding to a longitudinal translational temperature of $10.5~\si{\kelvin}$ \cite{miller1988a}. 

\begin{table}[t]
\caption{Beam and metastable parameters. }
\label{tab:params}
\begin{ruledtabular}
\footnotesize
\setlength{\tabcolsep}{4pt}
\begin{tabular}{lll}
quantity & value & source\\
\hline
molecules per pulse $N_\text{pulse}$ & $\num{1.1e16}$ & pressure rise\\
beam velocity $v$      & $2500~\si{\meter\per\second}$ & ToF peak\\
speed ratio $S$        & $8.4$                   & ToF width\\
$T_\parallel$          & $10.5~\si{\kelvin}$       & $mv^2/(2k_BS^2)$\\
$\Delta v_\text{FWHM}$ & $493~\si{\meter\per\second}$ & $1.665\,v/S$\\
\hline
gas beam divergence \footnotemark[1]     & \ang{12} FWHM & \cite{irimia2009a}\\
thermal cone            & \ang{11} FWHM          & from $T_\parallel$\\
metastable divergence   & \ang{16} FWHM          & quadrature sum\\
$\Omega_\text{skimmer}$ & $\num{1.13e-2}~\si{\steradian}$ & geometry\\
$\Omega_\text{MCP}$     & $\num{1.16e-3}~\si{\steradian}$ & geometry\\
skimmer transmission    & $20\,\%$ / $11\,\%$    & gas / metastable\\
MCP acceptance $\eta_\text{MCP}$         & $1.2\,\%$              & of the full beam\\
\hline
metastables per pulse $N_\text{det}$    & $\num{2.5e6}$ & from MCP \\
excitation fraction $f$ & $\num{1.9e-8}$         & $N_\text{det}/(\eta_\text{MCP}N_\text{pulse})$\\
$n(c\,^3\Pi_u^-)$ at laser\footnotemark[2] & $\num{7e5}~\si{\per\cubic\centi\meter}$ & $N_\text{det}/(\Omega_\text{MCP}\tau z^2v)$\\
\end{tabular}
\end{ruledtabular}
\footnotetext[1]{The divergence of the conical nozzle common to this valve family is taken from Ref.~\onlinecite{irimia2009a}.}
\footnotetext[2]{At this position $\tau=\SI{32}{\micro\second}$.}
\end{table}

\subsection{Metastable Yield}
The absolute metastable yield is estimated by taking typical two-stage (chevron stack) MCP gain characteristics \cite{hamamatsu2006mcp} at the applied supply voltage (potential difference over the MCP plates of approximately $1.3~\si{\kilo\volt}$), corresponding to an expected gain of approximately $10^5$.
An effort was made to calibrate the MCP gain by measuring single-particle counts, but these could not be distinguished from the noise floor at the low gains that were used.
Higher gains were not safely accessible without accelerating the degradation of the MCP because of the high signal levels caused by the fluorescence.
Furthermore, the detection efficiency of an MCP for neutral metastables is not well characterized in the literature; we adopt an estimate of $10\,~\si{\percent}$, motivated by the analogy that a metastable with $11.8\,~\si{\electronvolt}$ of internal energy deposits a comparable energy into the MCP surface upon impact as an ion with an equivalent kinetic energy.
Under these assumptions, the absolute particle number is solely an order-of-magnitude estimate, but the relative change in the number of measured particles is accurate.
The total charge associated with the current signal from the MCP was obtained by integrating under the signal curve and dividing by the $50~\si{\ohm}$ input impedance of the preamplifier to convert the volt-seconds measurement to coulombs.
The total metastable flux per shot at the MCP is estimated at $\num{2.0e6}$ total particles in the $V_\text{in}=80~\si{\volt}$ (the minimum for consistent ignition) to roughly $150~\si{\volt}$ range (see Fig. \ref{fig:yield_figs}(a)), rising to $\num{2.5e6}$ at $230~\si{\volt}$, the highest voltage for which the metastables were characterized.
The peak metastable flux stayed almost constant over the full range of tested voltages, and the increased total yield was mostly the result of metastables being produced over a longer time interval, as seen in Fig. \ref{fig:yield_figs}(b).

Converting the detected count into an excitation fraction requires the angular distribution of the beam, which was not measured here.
For the conical nozzle common to this valve family, we assume a beam divergence of $12^\circ$ FWHM \cite{irimia2009a}, with a Gaussian profile. 
The metastables are more divergent than the gas, because they stop colliding at the moment they are created, so they retain a transverse velocity spread corresponding to the measured $T_\parallel$. To obtain the angular FWHM for the metastable beam we add this thermal cone ($11^\circ$) in quadrature to the beam divergence given by the nozzle. 
The skimmer therefore transmits $11\,\%$ of the metastables against $20\,\%$ of the gas, and the MCP subtends $1.2\,\%$ of the full metastable beam.
The detected $\num{2.5e6}$ metastables correspond to $\num{2e8}$ metastables in the beam at the detector, and an excitation fraction of $\num{1.9e-8}$. We do not include a correction factor for the lifetime.  
Varying the assumed divergence over the reported range, and the shape of the angular profile, changes this fraction by $40\,\%$, which is small compared with the uncertainty in the detection efficiency.
The metastable density at the laser is independent of the angular distribution altogether, since the MCP measures the on-axis intensity directly.
Table \ref{tab:params} collects the beam and metastable parameters.

The REMPI H$_2^+$ signal (see subsection V.D) was found to increase as the pressure of the supplied H$_2$ gas was decreased.
Technical limitations of the gas-handling system made it unfeasible to record a detailed yield curve.
Observations during pressure optimization show an approximately fourfold yield increase as the stagnation pressure is reduced from $3.5$~bar to $1.2$~bar.

To test if other species contribute to the metastable signal, an Ar-H$_2$ mixture was used and the metastable signal on the MCP increased significantly. Ar metastables were detected by Rydberg spectroscopy. Under normal operating conditions and after purging the gas line, no Ar signal was found. A search for metastable H(2s) atoms by 1+1 REMPI via the 2s~$\rightarrow$~3p transition also resulted in no signal. The metastable signal is therefore believed to originate mainly from H$_2$. We cannot exclude contributions below our detection limit and the reported excitation fraction is an upper limit.

\subsection{Fluorescence}
\label{subsec:fluorescence}

The signal detected on the MCP during the discharge is vacuum-ultraviolet fluorescence, dominated by the dipole-allowed singlet systems of H$_2$ (the Lyman and the Werner bands) together with H Lyman-$\alpha$ from dissociative excitation.
The $a~^3\Sigma_g^+\rightarrow b~^3\Sigma_u^+$ continuum responsible for the visible plume of Fig. \ref{fig:discharge_cad}(c) lies in the near-ultraviolet and visible, outside the response of a bare MCP. The fluorescence thus monitors the singlet channel, not the triplet manifold from which the metastables originate. The total integrated signal increased monotonically with discharge voltage, by a factor of roughly $4.5$ from $V_\text{in}=85~\si{\volt}$ to $230~\si{\volt}$. A quantitative measurement was not attempted, as the strong initial signal almost certainly saturates the MCP even at the lowest usable gain.

The fluorescence, Fig. \ref{fig:yield_figs}(c), appears to consist of two signal components, a slow-varying baseline signal with fast spikes that we interpret as DBD ignition events. These can be separated in the experimental data by using an asymmetric least squares fit, with the constraint that the fit remains positive. More information and source code is provided in the Supplementary Material. The $B$ and $C$ states radiate in less than a nanosecond, so the emission follows the instantaneous electron flux and the spikes resolved in Fig. \ref{fig:yield_figs}(d) are individual microdischarges. Each event is short-lived because charge accumulating on the dielectric cancels the applied field locally and extinguishes it.

The isolated ignition peaks at $V_\text{in}=85~\si{\volt}$ and $230~\si{\volt}$ show clear differences. The initial ignition happens at a similar timing, but at $230~\si{\volt}$ the intensities and the frequency of the ignitions are significantly higher. Both observations are expected, since a higher applied field brings more sites above breakdown and lets site each transport more charge before the field collapses. The slowly varying component is not resolved into individual events.
It may be the unresolved superposition of weaker microdischarges, or an afterglow emission in which Lyman-$\alpha$ radiation is emitted following dissociative recombination for as long as ions and electrons persist.

Because the fluorescence reports the singlet channel, it is not a proxy for metastable production.
The singlet cross sections peak near $60~\si{\electronvolt}$ and fall only as $\ln E/E$, whereas the exchange-mediated triplet cross sections peak near threshold and fall as $E^{-3}$ \cite{janev2003a}, so a rise in mean electron energy raises the fluorescence considerably faster than the $c$-state excitation rate.

\subsection{Spectroscopic Analysis}
\begin{figure*}[htbp!]
    \centering
    \includegraphics[width=0.8\linewidth]{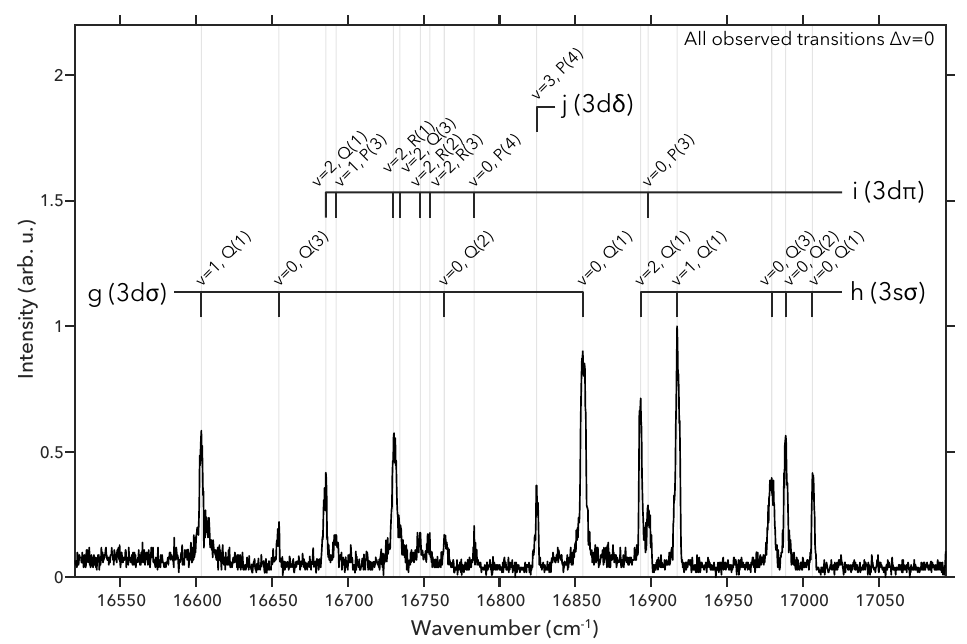}
    \caption{Spectrum of the n=3 \textit{gerade} complex measured via 1+1 REMPI from the c$^3\Pi_u^-$ state. The gray lines indicate the transition wavenumbers taken from Ref.~\onlinecite{jozefowski1994a}.}
    \label{fig:spectrum}
\end{figure*}
Air contamination from leaks in the gas handling system has been observed to give a metastable ToF signal an order of magnitude stronger than with pure H$_2$.
In these cases, H$_2$ acts as a carrier and accelerates the heavier metastable species via collisions in the supersonic expansion.
With specially prepared mixtures of Ar and H$_2$, we found that the inclusion of the heavier gas in small concentrations gives a significantly stronger metastable signal (by about an order of magnitude) and only causes a small delay in the ToF of the metastable peak.
Therefore, ToF measurements alone are insufficient to determine the composition of a measured metastable signal.

To confirm the composition of the measured metastables, a spectrum was measured using a pulsed dye laser.
Pulses with a duration of $5~\si{\nano\second}$ and an energy of approximately $1~\si{\milli\joule}$ in the wavelength range between $585~\si{\nano\meter}$ and $605~\si{\nano\meter}$ were obtained by pumping sulforhodamine B dye with the second harmonic a pulsed Nd:YAG laser at $532~\si{\nano\meter}$.
In this range, transitions can be driven from the c$^3\Pi_u^-$ state to the principal quantum number $n=3$, \textit{gerade} complex of states.
Ions are obtained via $1+1$ REMPI and the spectrum is shown in Fig.~\ref{fig:spectrum}. Assignments have been made using the tables provided by J\'ozefowski and co-workers \cite{jozefowski1994a}.

Transitions were driven with a DC electric field of $67~\si{\volt\per\centi\meter}$ along the beam direction.
The accelerated ions were energy-selectively guided to the electron multiplier (see Fig. \ref{fig:discharge_cad}, panel (d)).

The rotational temperature was estimated from the ratios of the line intensities of the $Q(1)/Q(3)$ and $R(1)/R(3)$ pairs, taking into account the linestrength ratios \cite{verdegay2025a} and after applying a correction based on the measured laser intensity curve.
The $v=0$ Q(3) and Q(1) transitions to the g ($3$d$\sigma$) state led to an estimate of the rotational temperature of $190~\si{\kelvin}$, and the $v=2$ R(1) and R(3) transitions to the i ($3$d$\pi$) state gave a value of $197~\si{\kelvin}$. The blended $h ~{}^3\Sigma_g^+ \leftarrow c ~{}^3\Pi_u^- \;(0,0)\; Q(3)$ line was not used for estimating the temperature. The obtained temperature is below the 230~K observed by Sprecher and coworkers \cite{sprecher2013a}. No lines corresponding to rotational states with $N>4$ have been observed.
The rotational temperature exceeds the longitudinal translational temperature of the beam by more than an order of magnitude.
This is characteristic of pure H$_2$, because of the large rotational constant in combination with ortho--para conversion being forbidden in collisions.
Because the rotational levels of H$_2$ are widely spaced, rotational relaxation is slow and about 300 collisions are needed at $300~\si{\kelvin}$ \cite{gallagher1974a}. Rotational cooling therefore stops within a few nozzle diameters, as the collision rate falls, while translational cooling continues. Measurements on pure H$_2$ jets at comparable stagnation conditions give a frozen rotational temperatures of $182~\si{\kelvin}$ \cite{gallagher1974a}.  Consequently, the present value is close to the expected cooling limit.

\section{Discussion}
Three observations help to clarify the mechanism that limits the metastable yield: (i) The monotonic improvement with decreasing stagnation pressure is consistent with collisional quenching as the likely dominant pressure-dependent loss. (ii) The weak dependence on supply voltage indicates that the additional excitation produced at high voltage is created where it cannot survive. (iii) The low absolute excitation fraction is set by neither, but by the branching within the triplet manifold.
The metastable c$^3\Pi_u^-$ and a$^3\Sigma_g^+$ states have an exceptionally large quenching cross-section upon collision with ground-state H$_2$ molecules, $\sigma_q \approx \num{7.5e-15}\,~\si{\centi\meter^2}$, as measured by Phelps and coworkers \cite{tischer1985a,wedding1988a}. This value corresponds roughly three times the gas-kinetic cross section defined by the Lennard-Jones diameter \cite{gallagher1974a}, so essentially every collision quenches.

Metastable H$_2$ molecules that reach the detector have not made any collisions since the moment they were created. Their velocity distribution is therefore frozen at that moment, and the measured speed ratio represents a snapshot of the local flow conditions upon metastable formation. The detected metastables are not a representative sample of those produced, but only the subset created late enough, and far enough from the nozzle that no further collision occurred.

\subsection{Excitation fraction}
The excitation fraction of $\num{1.9e-8}$ is low, but not unexpected for H$_2$ c$~^3\Pi_u^-$ metastable molecules.
Halfmann and coworkers report a production efficiency of about $10^{-7}$ for metastable helium \cite{halfmann2000a}, within an order of magnitude of the present value.
Luria and coworkers report $\num{6e-4}$ for metastable neon from a DBD of the same concept \cite{luria2009a}, four to five orders of magnitude higher.
This difference can be explained by: (i) the branching within the triplet manifold accounting for two orders of magnitude, as explained in the following section; (ii) collisional quenching in H$_2$ accounting for a further factor of three, and (iii) the possibility that the detection efficiency of metastable H$_2$ on the MCP is even lower than the assumed $10\,\%$.

The uncertainty of our value is dominated by the assumed MCP detection efficiency and by the gas-load calibration (assumed to be correct within one order of magnitude). 

\subsection{Branching within the triplet manifold}
Reaching the triplet manifold is not the same as producing a detectable metastable. Only $10$--$18\,\%$ of the triplet excitation products result in $c~^3\Pi_u$ state population for mean electron energies between $3$ and $10~\si{\electronvolt}$, the rest going predominantly into the repulsive $b~^3\Sigma_u^+$ state \cite{scarlett2021a}. Half of that is the promptly predissociating $^3\Pi_u^+$ component. Based on the Franck-Condon factors, only $18\,\%$ of the $c~^3\Pi_u$ state population is in $v'=0$, with the distribution peaking at $v'=1$. Including the radiative decay of all levels during the flight to the detector, only about $1\,\%$ of the triplet excitation leads to the formation of a metastable molecule. 

The vibrational distribution is directly visible in the spectrum. Transitions from $v'=0$, $1$ and $2$ appear with comparable intensity, which at first sight conflicts with the order-of-magnitude shorter lifetime of $v'\geq1$. The two effects nearly cancel. The nascent population of $v'=1$ exceeds that of $v'=0$ by $37\,\%$, while its survival over the $80~\si{\micro\second}$ flight to the laser is lower by a factor of two, given $\tau(v'=0) \approx 1~\si{\milli\second}$ and $\tau(v'\geq1) \approx 100~\si{\micro\second}$ \cite{astashkevich2015a}. The predicted intensities relative to $v'=0$ are $0.67$ for $v'=1$ and $0.57$ for $v'=2$. The observation therefore confirms both the calculated branching and the assumed lifetime, neither of which was adjusted.

\subsection{Pressure dependence}
In a supersonic expansion the local number density scales linearly with the stagnation pressure at every point in the flow, since the Mach number profile is set by the geometry of the expansion alone. The two-body quenching rate $R_q = n\,\sigma_q\,\bar{v}_\text{rel}$ is therefore proportional to $P_0$, as is the quenching probability integrated along the trajectory. The number of metastables created also scales with density, and therefore the detected yield follows $Y \propto P_0\exp(-aP_0)$, in which the prefactor represents the excitation events and the exponential the fraction that survives.
The observed fourfold increase on reducing $P_0$ from $3.5$ to $1.2$~bar gives $a \approx 1.1~\si{\per\bar}$ and a survival probability of about $30\,\%$ at the operating point. Quenching therefore reduces the metastable density by a factor of three, not the many orders of magnitude that $\sigma_q$ alone might suggest. The yield $Y(P_0)$ is maximal at $aP_0 = 1$, implying that the source operates within a few per cent of the best stagnation pressure.
The model is consistent with our observations but has not been tested independently.

\subsection{Voltage dependence}
The total integrated fluorescence signal increases by a factor of roughly $4.5$ between $V_\text{in} = 85\,~\si{\volt}$ and $230\,~\si{\volt}$, indicating that the discharge produces a substantially larger excited-state population at higher voltages. Yet over the same range the metastable yield increases far less, with the peak flux remaining nearly constant and the total yield growing only because the discharge is sustained over a longer time interval at higher supply voltage.
Specifically, the $85\,~\si{\volt}$ case produces a peak metastable flux that is only approximately $7\,~\si{\percent}$ lower than at $230\,~\si{\volt}$, despite the discharge being far less intense.

Three mechanisms can explain these observations.

(i.) The additional metastables are created where they cannot escape. A higher applied voltage permits ignition at higher gas density, earlier in the gas pulse and closer to the nozzle. The extra discharge activity at $230\,~\si{\volt}$ therefore occurs preferentially in the region where the survival probability is lowest, and contributes little to the detected flux. 

(ii.) As discussed in Sec.~\ref{subsec:fluorescence}, the fluorescence corresponds to the singlet channel, for which the rate coefficients rise considerably faster with electron energy than for the exchange-mediated triplet channels. 

(iii.) The lowest vibrational levels of the  $c~^3\Pi_u$ and $a~^3\Sigma_g^+$ states are separated by only $0.017~\si{\electronvolt}$, and the cross section for electron-impact transfer between them is correspondingly large, peaking at $\num{1.3e-12}~\si{\centi\meter\squared}$ \cite{janev2003a}. Since $a~^3\Sigma_g^+$ radiates into the repulsive $b~^3\Sigma_u^+$ state within $10~\si{\nano\second}$, the transfer destroys the triplet population. Averaged over a Maxwellian electron energy distribution at $3~\si{\electronvolt}$ the rate coefficient is $\num{4.9e-6}~\si{\centi\meter\cubed\per\second}$, exceeding the quenching rate coefficient for ground-state H$_2$ by three orders of magnitude.  Consequently, electron impact dominates once the ionization fraction exceeds $\num{4e-4}$. 

The ground-state H atoms produced by the dissociative triplet channels are unlikely to contribute. Even though their density also grows with discharge intensity, scaling the H-atom density reported for a comparable DBD source \cite{scheidegger2022a} to our beam gives a dissociated fraction below $10^{-4}$. H atoms would need a rate coefficient four orders of magnitude above that of H$_2$ to play a role in the quenching process.

\section{Conclusion}
A pulsed supersonic beam source of metastable H$_2$ molecules in the $c~^3\Pi_u^-$ state is produced by a DBD integrated in the nozzle of a piezo valve.
The off-axis detection of H$_2^+$ ions produced by 1+1 REMPI of the $c~^3\Pi_u^-$ state via the $n=3$ gerade complex confirms that the detected metastable signal originates from H$_2$ rather than contaminant species.
The measured total metastable flux per shot ranges from approximately $\num{2.0e6}$ at the ignition threshold to $\num{2.5e6}$ at $230\,~\si{\volt}$, corresponding to $\num{2e8}$ metastables created per pulse and an excitation fraction of $\num{1.9e-8}$.
This is within an order of magnitude of the efficiency reported for metastable helium from a filament-seeded pulsed discharge \cite{halfmann2000a}, despite the spin-forbidden excitation, the dissociative character of the entire triplet manifold, and the absence of an inert carrier gas.

The monotonic improvement of the yield with decreasing stagnation pressure is consistent with collisional quenching as the dominant pressure-dependent loss, with a survival probability of about $30\,\%$ at the operating point.
The weak dependence on the supply voltage indicates instead that the additional excitation produced at higher voltage is created at densities where the metastable molecules do not survive.
The absolute scale of the yield is set by neither, but by the branching within the triplet manifold, which delivers of order one per cent of the triplet excitation to the detector as a metastable molecule.
The observation of comparable signal from $v'=0$, $1$ and $2$ is consistent with the calculated vibrational branching and with the shorter lifetime of the excited vibrational levels.

\section*{Supplementary Material}
The supplementary material provides PCB layout details (including recommended improvements to our current design), a description of the process of measuring and fitting DBD drive output waveforms with an equivalent circuit, and an explanation of the asymmetric least-squares procedure used to subtract the slow-changing baseline of the fluorescence signal to obtain Fig. \ref{fig:yield_figs} (d).

\section*{Acknowledgments}
We thank H. Schmutz for providing circuit schematics of the DBD driver used in Ref.~\onlinecite{scheidegger2022a} and J. Slijper and T. Warmerdam-Brouwer for technical support. We acknowledge helpful discussions with J. Agner, T. Schuitevoerder, A. D. Ochoa, W. van der Zande and M. H. M. Janssen and we thank F. Merkt for carefully reading the manuscript and for many useful suggestions.

\section*{Author Declarations}
\subsection*{Conflict of Interest}
The authors have no conflict of interest to disclose.

\subsection{Author Contribution}
\textbf{D.~Knapp}: Investigation (lead); Formal analysis (equal); Software (lead); Visualization (lead); Writing -- original draft (equal).
\textbf{M.~Beyer}: Conceptualization (lead); Supervision (lead); Formal analysis (equal); Writing -- review \& editing (lead).

\subsection*{Data Availability}
The CAD file for the alumina nozzle and the design files for the DBD PCB are provided.  SPICE model files and Python code used to perform least-squares fits are also available, along with the raw data from oscilloscope measurements of the voltage output.  Other raw data is available from the corresponding author upon reasonable request. Zenodo DOI: 10.5281/zenodo.23505172.

\section*{Funding}
This research was supported by the Dutch Research Council (NWO) through an ENW-M2 (OCENW.M.21.141, DOI: 10.61686/RHUGS31235) grant.

\section*{References}
%

%
%

\appendix

\section{Circuit Description}\label{subsec:circuit_description}
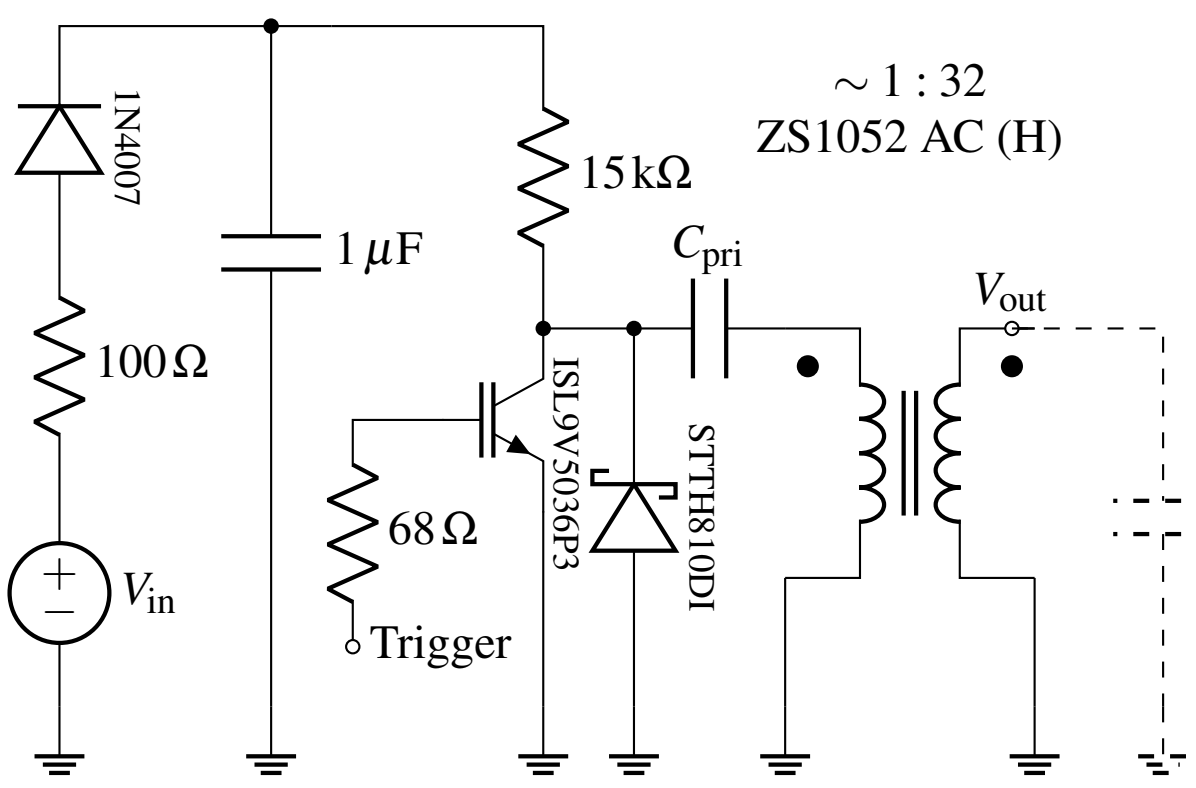
\begin{figure}[htbp!]
\begin{circuitikz}[american]
  \ctikzset{bipoles/length=1.1cm}

  \draw (0,0) node[ground]{} to[V, invert, l_=$V_{\text{in}}$] (0,1.5)
        to[R, l_=$100\,\Omega$] (0,3)
        to[D] (0,4.5) -- (1.8,4.5);
        \node[rotate=-90, font=\scriptsize] at (0.45, 3.7) {1N4007};

  \draw (1.4,4.5) to[C, l=$1\,\mu\text{F}$, *-] (1.4,1.5) -- (1.4,0) node[ground]{};

  \draw (1.8,4.5) -- (3.2,4.5) to[R, l=$15\,\text{k}\Omega$, -*] (3.2,2.5);

  \draw (3.2,2.5) node[nigbt, anchor=C] (igbt) {};
  \draw (igbt.E) -- (3.2,0) node[ground]{};
  
  \node[rotate=-90, font=\scriptsize] at (3.35, 1.6) {ISL9V5036P3};

  \draw (igbt.G) -- ++(-0.6,0) to[R, l=$68\,\Omega$] ++(0,-1.5)
        node[ocirc]{} node[right]{Trigger};

  \draw (igbt.C) -- (3.8,2.5) to[C, l=$C_\text{pri}$] (4.8,2.5);

  \draw (3.8,0) node[ground]{} to[sD, -*] (3.8,2.5);
  \node[rotate=-90, font=\scriptsize] at (4.25, 1.25) {STTH810DI};

  \node[transformer core, anchor=A1] (T) at (4.8,2.5) {};
  
  \node[align=center, above] at ([yshift=1cm]T.north) {$\sim1:32$ \\ ZS1052 AC (H)};

  \node[circ, scale=1.5] at ([xshift=0.15cm, yshift=-0.25cm]T.A1) {};
  \node[circ, scale=1.5] at ([xshift=-0.15cm, yshift=-0.25cm]T.B1) {};

  \draw (T.A2) -- (T.A2 |- 0,0) node[ground]{};
  \draw (T.B2) -- (T.B2 |- 0,0) node[ground]{};

  \draw (T.B1) -- (6.3,2.5) node[ocirc]{} node[above]{$V_{\text{out}}$};

  \draw[dashed] (6.3,2.5) -- (7.3,2.5) to[C] (7.3,0) node[ground]{};
\end{circuitikz}
\caption{Schematic of the in-vacuum DBD driver circuit. The capacitive load represented by the output electrode is indicated using dashed lines.}\label{fig:Schem}
\end{figure}

The DBD is driven by high-voltage pulses from a capacitive discharge ignition circuit, shown schematically in Fig. \ref{fig:Schem}.
These circuits are well-established for providing the increased potential necessary to initially ignite a gas discharge tube, and have been demonstrated for driving DBDs \cite{tang2022a}.
A resonant LC circuit is created by $C_\text{pri}$ and the primary of the autotransformer (Excelitas Technologies, ZS1052 AC (H)).
The value of $C_\text{pri}$ should be tuned as described below.
The results presented in the main text were measured with $C_\text{pri}=330~\si{\nano\farad}$, and the primary had a measured inductance of $20~\si{\micro\henry}$.
In isolation, this would suggest a resonant frequency of $1/2\pi\sqrt{L_\text{pri}C_\text{pri}}\approx62~\si{\kilo\hertz}$, but the strong coupling to the secondary (measured inductance of $20~\si{\milli\henry}$) causes a shifted resonance frequency of $56~\si{\kilo\hertz}$.
The secondary is unloaded except for the capacitance of the DBD electrode and the parasitic parallel capacitance of the secondary windings, resonant at approximately $500~\si{\kilo\hertz}$.
The output voltage $V_\text{out}$ is a superposition of two decaying oscillations at these two characteristic frequencies.

Before triggering, $C_\text{pri}$ is charged to voltage $V_\text{in}$ (to within the forward voltage of the input reverse polarity protection diode).
When the circuit is triggered, the IGBT (OnSemi ISL9V5036P3) conducts, effectively tying the collector node to ground, completing the primary LC circuit loop.
This creates a voltage of $-V_\text{pri}$ at the primary of the autotransformer, acting as the impulse that excites the coupled primary and secondary resonances.

The trigger is a $5~\si{\volt}$ TTL pulse from a delay generator.
A conservatively long duration of $10~\si{\milli\second}$ was chosen to allow oscillations to damp out before the connection to ground is opened.
Opening the connection (turning off the trigger pulse) while the IGBT is conducting current driven by the resonating primary, risks a high voltage pulse, potentially destroying the IGBT.
The inductive kickback protection diode in parallel with the IGBT (STMicroelectronics STTH810DI) is intended to protect from this failure mode.
The time constant for recharging $C_\text{pri}$ is $5~\si{\milli\second}$.
Consequently, there is ample time for the circuit to reset during the $10~\si{\hertz}$ repetition rate.
The circuit was tested under experimental conditions for $10~\si{\minute}$ with $V_\text{in}=250~\si{\volt}$ without failure.
The failure point of the circuit has not been tested.

At the $85~\si{\volt}$ operating point in Fig. \ref{fig:yield_figs}, a peak voltage of about $3.5~\si{\kilo\volt}$ was delivered to the DBD electrode, assuming that the circuit behaves linearly, and the voltage is not lowered by (dis)charging of the dielectric surface by the DBD microdischarges.
This estimated value should be treated as an upper bound for the potential that appears in the actual experiment.
The minimum voltage for consistent DBD ignition, $80~\si{\volt}$, corresponds to about $3.3~\si{\kilo\volt}$, which can be treated as approximately the peak potential that the DBD electrode reaches before microdischarges are ignited.
The total energy delivered by the DBD pulse is approximately $1.2~\si{\milli\joule}$ for the $85~\si{\volt}$ input voltage operating point, and $8.7~\si{\milli\joule}$ at $230~\si{\volt}$, under the assumption that the entirety of the energy stored in the primary pre-ignition is dissipated into the gas.
The peak current reaching the DBD electrode during the course of the $500~\si{\kilo\hertz}$ ringing of the secondary is estimated to be on the order of about $30~\si{\milli\ampere}$.
If the estimated peak voltage of $3.5~\si{\kilo\volt}$ were to be discharged from the DBD electrode (fitted effective capacitance of approximately $2.5~\si{\pico\farad}$) during the FWHM duration ($15~\si{\nano\second}$) of the initial sharp fluorescence peak in Fig. \ref{fig:yield_figs}(c), the instantaneous current flowing into the plasma is estimated to be approximately $600~\si{\milli\ampere}$.

\begin{figure}
    \centering
    \includegraphics[width=0.8\linewidth]{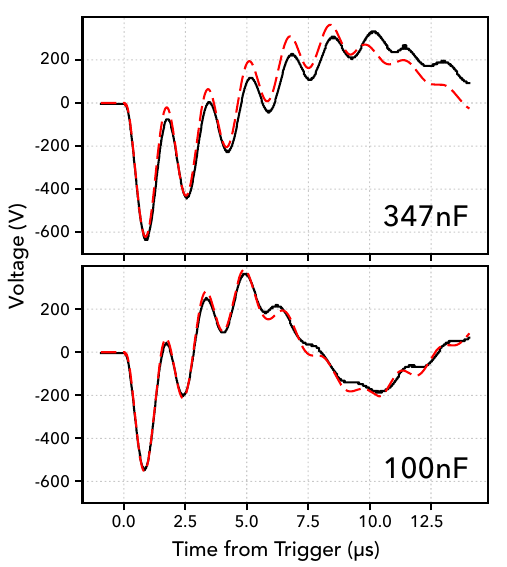}
    \caption{Comparison of least-squares fitted SPICE model (red dashed line) and experimental data (black solid line), extrapolated to $C_{\text{pri}}$ values that were not used as input for the fit.}
    \label{fig:spice-model}
\end{figure}
The circuit is a capacitive discharge ignition type, commonly seen for triggering of gas discharge lamps.
Topologically, it bears similarities to the `Tesla Transformer' circuit analyzed (in the lossless case) by Cook and Reginato \cite{cook1979b}, where an analytical model of the circuit is developed.
It differs from a Tesla Coil in the sense that the energy exchange between the primary and secondary is non-resonant by design.
It differs from a Tesla Coil in the sense that the energy exchange between the primary and secondary is non-resonant by design.
The sharp initial transient triggered by the IGBT excites oscillation at two distinct resonance frequencies, and the output is the decaying superposition of them.
The advantage of the non-resonant coupling is that the primary can use a relatively large-valued capacitor $C_\text{pri}$ to couple energy into the output waveform, resulting in a lower primary resonance frequency, while the secondary resonates quickly providing a higher output slew rate that promotes frequent microdischarge ignition.

Fig. \ref{fig:spice-model} shows the SPICE (Simulation Program with Integrated Circuit Emphasis) model being used to predict the output waveform for two different values of $C_{\text{pri}}$ not included in the input to the SPICE model fit, demonstrating the fidelity of the model to the actual implemented circuit.
This characterization supports reproduction, adjustment, and part substitution in the source electronics, especially given that the parasitics of the transformer are not well-characterized in the documentation provided by the manufacturer.
Future investigations could benefit from detailed modeling of the coupling between the plasma and the equivalent circuit presented in this work.

\begin{figure}[htbp!]
    \centering
    \includegraphics[width=0.8\linewidth]{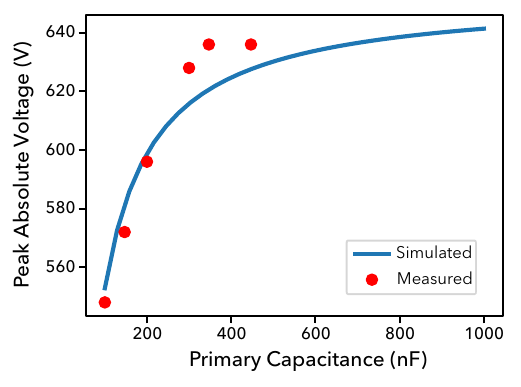}
    \caption{Comparison of simulated and measured peak output voltages for various capacitances $C_\text{pri}$ with input voltage $V_\text{in}=15\si{\volt}$.}
    \label{fig:peak_volt_curve}
\end{figure}
The primary capacitor $C_\text{pri}$ plays two distinct roles.
It stores the energy, $C_\text{pri}V_\text{in}^2/2$, that is released into the coupled oscillator system, and it sets the resonance frequency of the primary.
In a perfectly linear circuit, the output voltage would increase roughly as $\sqrt{C_\text{pri}}$, which can be seen by assuming an efficiency constant $\eta$ that represents the maximum fraction of the total energy that is transferred to the secondary at any given moment.
Therefore,
\begin{equation}
    \begin{split}
        \frac{1}{2}\eta C_\text{pri}V_\text{in}^2&=\frac{1}{2}C_\text{sec}V_\text{out, max}^2\\
        \frac{V_\text{out, max}}{V_\text{in}}&=\sqrt{\eta\frac{C_\text{pri}}{C_\text{sec}}}\ .
    \end{split}
\end{equation}
In practice, as larger capacitors are used, non-linear losses from transformer core saturation are expected to cause even more efficiency degradation at high peak currents.
Consequently, maximizing $C_\text{pri}$ does not yield monotonically better performance.
Furthermore, a fundamental trade-off exists between stored energy and resonant frequency.
While a smaller $C_\text{pri}$ reduces the total energy per pulse, it increases the resonant frequency of the primary circuit.

Fig. \ref{fig:peak_volt_curve} shows measured and simulated peak voltages for various primary capacitances.
Rather than programming the measured ESR values for each tested capacitance, the curve was produced by assuming that $\text{ESR}(C_\text{pri})=\left(1\si{\ohm}\right)(100\si{\nano\farad})/C_\text{pri}$.
This is the primary cause of the scatter of the data points (the larger tested capacitances were made by placing up to four capacitors in parallel, causing the effective ESR to be lower and the efficiency to be higher), but it gives an approximate trend that can be extrapolated to larger capacitances.
Even as the energy introduced into the circuit increases linearly with $C_\text{pri}$, the maximum output voltage sees diminishing returns.

\end{document}


\title{\texorpdfstring{Supplementary Material for\\
Compact Dielectric Barrier Discharge Source for a Pulsed Supersonic Beam of Metastable Triplet H$_2$}{Supplementary Material for Compact Dielectric Barrier Discharge Source for a Pulsed Supersonic Beam of Metastable Triplet H2}}
\author{Daniel Y. Knapp}
\email{d.y.knapp@vu.nl}
\affiliation{Department of Physics and Astronomy, LaserLaB, Vrije Universiteit Amsterdam, De Boelelaan 1100, 1081 HZ Amsterdam, The Netherlands}
\author{Maximilian Beyer}
\email{m.beyer@vu.nl}
\affiliation{Department of Physics and Astronomy, LaserLaB, Vrije Universiteit Amsterdam, De Boelelaan 1100, 1081 HZ Amsterdam, The Netherlands}

\renewcommand{\thesection}{S\arabic{section}}
\renewcommand{\thesubsection}{\thesection.\arabic{subsection}}
\renewcommand{\thesubsubsection}{\thesubsection.\arabic{subsubsection}}
\renewcommand{\thefigure}{S\arabic{figure}}
\renewcommand{\thetable}{S\arabic{table}}
\renewcommand{\theequation}{S\arabic{equation}}
\renewcommand{\thepage}{S\arabic{page}}
\maketitle
\pagestyle{plain}
\thispagestyle{plain}
\raggedbottom

\noindent
This supplementary material documents the details of the in-vacuum PCB implementation (\ref{sec:pcb}), SPICE (Simulation Program with Integrated Circuit Emphasis) fitting details (\ref{sec:spice}), and the assymetric least-squares fluorescence baseline subtraction (\ref{sec:fluor}).

\section{PCB Implementation}\label{sec:pcb}
\begin{figure*}\label{fig:pcb}
    \centering\includegraphics[width=0.75\linewidth]{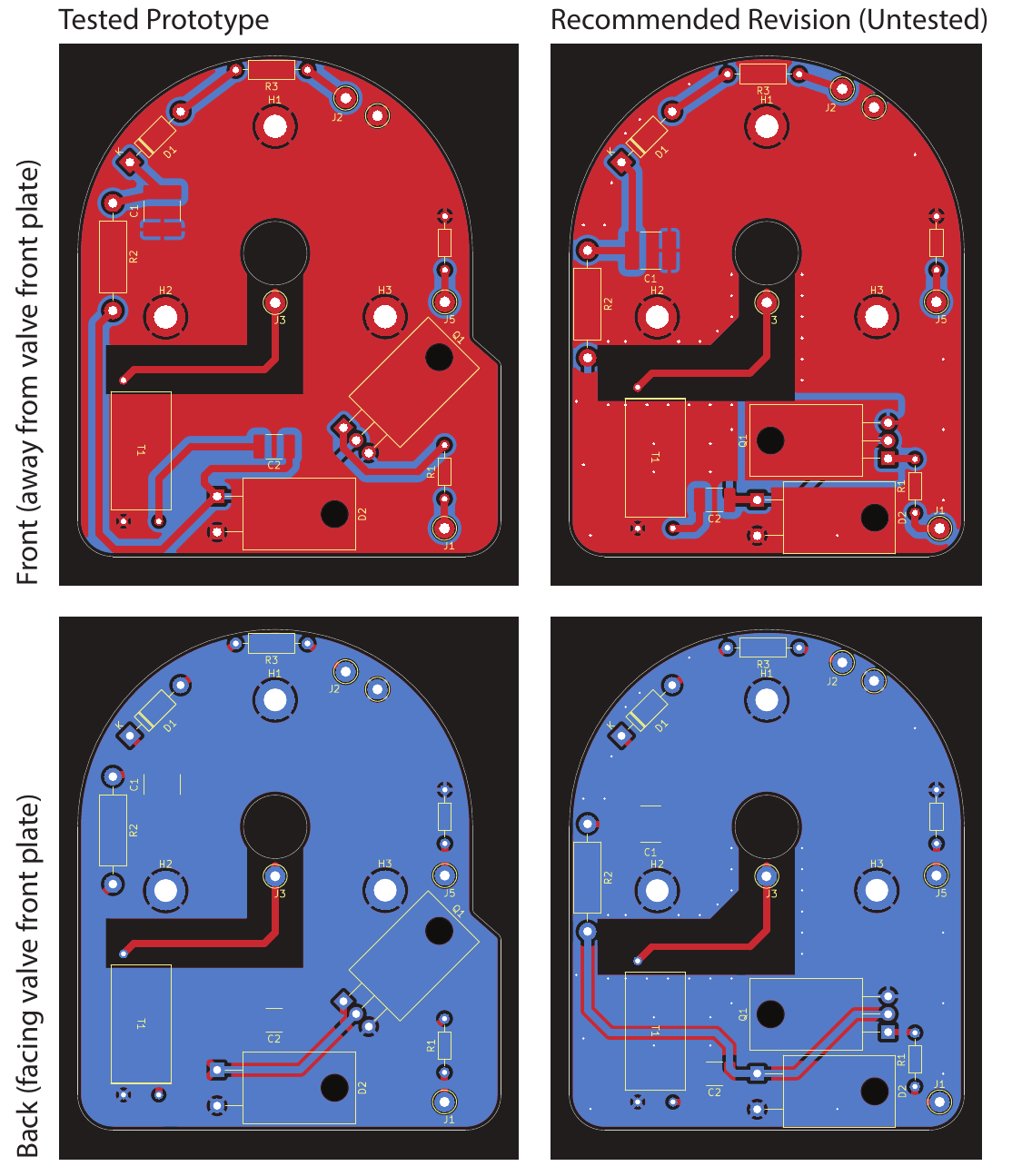}
    \caption{Illustration of PCB layout for tested version (left) and recommended (but untested) revision (right).  Layers are F.Cu (red), B.Cu (blue), Edge.Cuts (white lines), drilling (white filled circles), and silkscreen (ivory).  The silkscreen layer is omitted in the final design files sent to the manufacturer but included in this figure to illustrate component placement.  The netlist is identical for both shown designs; only component placement and routing is changed.}
    \alttext{Four renderings of the circuit board layout. The left column shows the tested prototype and the right column the recommended but untested revision; the top row shows the front copper layer in red and the bottom row the back copper layer in blue. Both boards have a rounded outline, three mounting holes and a central hole for the nozzle. The revision places the components more compactly with shorter connections.}
\end{figure*}
The PCB tested in this work (fig. \ref{fig:pcb}, left-hand side) was laid out as a rapid prototype under time constraint.
All of the reported results were obtained with it, and to our knowledge it is completely functional with no performance limitations related to the layout.
However, the routing and component placement is not optimized.
For readers that wish to reproduce this design, we provided a re-routed version (fig. \ref{fig:pcb}, right-hand side) with several improvements without any modifications to the netlist.
These improvements include:
\begin{enumerate}
    \item Reduced distance between gate isolation resistor R1 and the Q1 base, to reduce the risk of parasitic (and potentially damaging) oscillations.
    \item A filled zone that allows Q1 to be bolted to the PCB (not possible in the prototype design).
    \item Cleaner routing between Q1, D2, C2, and T1.
\end{enumerate}
However, we emphasize that the revised layout has not been produced or tested, and we make no claims about its performance.

The tested PCB was produced with $1.6~\si{\milli\meter}$ thickness FR-4 substrate with a $1~\text{oz}/\text{ft}^2$ copper layer and HASL coating.
For future production runs, we recommend ENIG coating for better vacuum compatibility.
Solder mask is excluded over the whole board so that the underlying conductor is exposed, for lower outgassing rates.

The PCB is mounted to the Amsterdam valve front plate using three M2.5 screws, through clearance holes H1-H3.
The valve itself supplies a ground connection to the PCB through these screws, so the additional wired ground connection is redundant.
The positive supply voltage input is J2.
The TTL pulse to trigger the DBD is applied to J1.
J5 is primarily intended for an NTC thermistor or similar, to monitor the valve temperature, since the Amsterdam piezo valve's gas output is highly temperature-dependent.
All of the data presented in this study were collected at room temperature.
The secondary side of T1, where high voltages are present, is protected from surface flashover by including a large creepage distance.
In practice, the PCB should be cleaned thoroughly before being introduced to the vacuum chamber, so flashover facilitated by surface contaminants is unlikely.
The loop of wire that acts as the DBD electrode should be pre-bent, test-fitted with the valve body, and only then soldered, to ensure that the ceramic nozzle is correctly aligned.

Initial cleaning was carried out using Chemtronics Flux-Off solvent and a toothbrush to remove flux residue.
The PCB was then rinsed and placed in a Petri dish with the components facing upward.
The Petri was filled with enough isopropanol to barely submerge the PCB, with the components left unsubmerged, out of concern that especially the transformer may absorb solvent if fully submerged.
The Petri dish was placed in an ultrasonic bath for five minutes, and then the same process was repeated using water.
A lint-free wipe was wetted with isopropanol and used to manually wipe the surfaces of the components.
Rigorous testing of the outgassing performance of the PCB has not been performed, but the chamber reaches a pressure of $\num{9.8e-8}\si{\milli\bar}$ when the Amsterdam valve is emptied, in line with other similar chambers in our laboratory that are sealed with viton O-rings and pumped down with the same model of turbomolecular pump.
The average pressure during operation of the valve is more than an order of magnitude higher, allaying concerns that significant contamination may be caused by outgassing.

\section{SPICE Fitting}\label{sec:spice}
\begin{figure*}[htbp!]
\centering
\resizebox{\textwidth}{!}{%
\begin{circuitikz}[voltage dir=RP]
\draw (0,0) -- (2.5,0)
      (0,0) to[american voltage source,invert,l_={$V_3=-0.45\,\mathrm{V}$}] (0,-1.65)
            -- (0,-2.12)
            to[american voltage source,invert,l_={$V_1=\textcolor{blue}{\{Vin\}}$}] (0,-3.77)
            node[ground]{}
      (2,0) to[C,l={$C_1=1\,\mu\mathrm{F}$}] (2,-2.0)
            -- (2,-3.77) node[ground]{}
      (2.5,0) to[R,l={$R_2=15\,\mathrm{k}\Omega$}] (4,0)
            -- (6,0);
\draw (6,0) -- (6,-0.55)
            to[american voltage source,invert,l={$V_4=\textcolor{blue}{\{VCE\}}$}] (6,-2.2)
            -- (6,-2.75)
            to[D,l_={$D_2$ \texttt{D\_IGBT}}] (6,-4.0)
            -- (6,-4.55)
            to[nos] (6,-6.3)
            -- (6,-6.75) node[ground]{};
\draw (6,-0.55) -- (8.25,-0.55) -- (8.25,-2.9);
\draw (8.25,-4.15) to[D,l={$D_3$ \texttt{D\_REV}}] (8.25,-2.9);
\draw (8.25,-4.15) -- (8.25,-5.1) node[ground]{};
\draw (4.75,-4.76) -- (5.43,-4.76)
      (4.75,-5.72) -- (5.43,-5.72);
\draw[fill=white] (5.43,-4.76) circle (0.06);
\draw[fill=white] (5.43,-5.72) circle (0.06);
\node[anchor=east] at (5.41,-4.56) {$-$};
\node[anchor=east] at (5.41,-5.92) {$+$};
\draw[thin,gray,rounded corners=3pt] (5.31,-4.40) rectangle (6.60,-6.48);
\node[anchor=west] at (6.65,-5.90) {$S_1$ \texttt{S\_IGBT}};
\draw (4.75,-4.76) -- (4.1,-4.76) -- (4.1,-5.08) node[ground]{}
      (4.75,-5.72) -- (4.75,-7.97)
            to[american voltage source,invert,l_={$V_2$}] (4.75,-9.62) node[ground]{};
\draw (6,0) -- (9.05,0)
      (9.05,0) to[C,l={$C_2$},a={$\textcolor{blue}{\{C2\}}$}] (10.25,0)
      (10.25,0) to[R,l_={$R_3=\textcolor{blue}{\{R3\}}$}] (11.75,0)
      (11.75,0) -- (12.3,0)
      (12.3,0) -- (14.5,0);
\node[ocirc, scale=0.75] at (12.3,0.55) {};
\draw (12.3,0) -- (12.3,0.55);
\node[above] at (12.3,0.58) {$V_{\rm pri}$};
\draw (12.3,0) to[R,l_={$R_4=\textcolor{blue}{\{R4\}}$}] (12.3,-1.65)
      (12.3,-1.65) to[L,l_={$L_4=\textcolor{blue}{\{L4\}}$}] (12.3,-3.35)
      (12.3,-3.35) to[R,l_={$R_7=1\,\mathrm{m}\Omega$}] (12.3,-5.05)
      (12.3,-5.05) -- (12.3,-5.58) -- (11.25,-5.58)
      (11.25,-5.58) to[L,l_={$L_3=\textcolor{blue}{\{L3\}}$}] (11.25,-7.33)
      (11.25,-7.33) -- (11.25,-9.53)
      (12.3,-5.58) -- (13.55,-5.58)
      (13.55,-5.58) to[L,l={$L_1=1\,\mathrm{mH}$}] (13.55,-7.33)
      (13.55,-7.33) to[R,l_={$R_9=1\,\mathrm{m}\Omega$}] (13.55,-9.08)
      (13.55,-9.08) -- (13.55,-9.53)
      (11.25,-9.53) -- (13.55,-9.53);
\draw (14.5,0) to[C,l={$C_4$},a={$\textcolor{blue}{\{C4\}}$}] (15.75,0)
      (15.75,0) -- (17.15,0);
\draw (17.15,0) to[R,l_={$R_5=\textcolor{blue}{\{R5\}}$}] (17.15,-1.65)
      (17.15,-1.65) to[L,l_={$L_5=\textcolor{blue}{\{L5\}}$}] (17.15,-3.35)
      (17.15,-3.35) to[R,l_={$R_8=1\,\mathrm{m}\Omega$}] (17.15,-5.05)
      (17.15,-5.05) -- (17.15,-5.58)
      (17.15,-5.58) to[L,l_={$L_2=1\,\mathrm{H}$}] (17.15,-7.33)
      (17.15,-7.33) to[R,l={$R_{10}=1\,\mathrm{m}\Omega$}] (17.15,-9.08)
      (17.15,-9.08) -- (17.15,-9.53)
      (13.55,-9.53) -- (17.15,-9.53);
\draw (15.45,-9.53) -- (15.45,-9.83) node[ground]{};
\fill (13.25,-7.02) circle (1.4pt);
\fill (16.85,-7.02) circle (1.4pt);
\node[align=center, font=\scriptsize] at (15.37,-8.1) {\texttt{K L1 L2 0.999}};
\draw (17.15,0) -- (18.9,0) -- (21.3,0) -- (23.4,0) -- (24.1,0);
\draw (17.15,-1.65) -- (18.9,-1.65)
      (18.9,-1.65) to[R,l={$R_6=\textcolor{blue}{\{R6\}}$}] (18.9,-3.35)
      (18.9,-3.35) -- (18.9,-5.58)
      (18.9,-5.58) to[C,l={$C_3=\textcolor{blue}{\{C3\}}$}] (18.9,-6.83)
      (18.9,-6.83) -- (18.9,-9.53) -- (17.15,-9.53);
\draw (21.3,0) -- (21.3,-2.85)
      (21.3,-2.85) to[C,l_={$C_5=\textcolor{blue}{\{C5\}}$}] (21.3,-4.1)
      (21.3,-4.1) -- (21.3,-5.1) node[ground]{};
\draw (23.4,0) -- (23.4,-2.85)
      (23.4,-2.85) to[R,l={$R_1=\textcolor{blue}{\{R1\}}$}] (23.4,-4.55)
      (23.4,-4.55) -- (23.4,-5.1) node[ground]{};
\node[ocirc,scale=0.8] at (24.1,0) {};
\node[right] at (24.15,0) {$V_{\rm out}$};
\foreach \x/\y in {
  2/0,6/0,6/-0.55,12.3/0,12.3/-5.58,
  13.55/-9.53,15.45/-9.53,17.15/0,17.15/-1.65,
  17.15/-9.53,18.9/-9.53,21.3/0,23.4/0}
  \fill (\x,\y) circle (1.1pt);
\node[anchor=west,font=\footnotesize] at (0,-10.6)
  {$V_2$: \texttt{PULSE(0 5 \textcolor{blue}{\{T\_ON\}} 50n 50n 1m 0.1)}};
\node[anchor=west,font=\footnotesize] at (0,-11.3)
  {\texttt{D\_IGBT: IS=10n N=2 RS=\textcolor{blue}{\{R\_FOR\}} CJO=100p VJ=0.7 M=0.33 TT=50n}};
\node[anchor=west,font=\footnotesize] at (0,-11.95)
  {\texttt{D\_REV: IS=10n N=2 RS=\textcolor{blue}{\{R\_REV\}} CJO=100p VJ=0.7 M=0.33 TT=\textcolor{blue}{\{TT\_REV\}}}};
\node[anchor=west,font=\footnotesize] at (0,-12.6)
  {\texttt{S\_IGBT: Vt=2.5 Ron=100m Roff=100Meg}};
\end{circuitikz}%
}
\caption{Equivalent circuit model used for SPICE fitting runs. Blue values denote fitted parameters. The parameters $C_2$, $R_3$, and $T_{\rm ON}$ were fitted separately for each measurement; the other fitted parameters were shared across measurements.}
\alttext{Equivalent circuit used for the SPICE fits. It contains the supply with a fixed diode offset, the charging network, the primary capacitor with its series resistance, the IGBT modelled as a switch with diodes, a transformer modelled by ideal coupled inductors with added leakage and magnetizing inductances, winding resistances and parasitic capacitances, and the electrode load as a capacitor in parallel with a resistor. Fitted parameters are marked in blue.}
\label{fig:equiv}
\end{figure*}
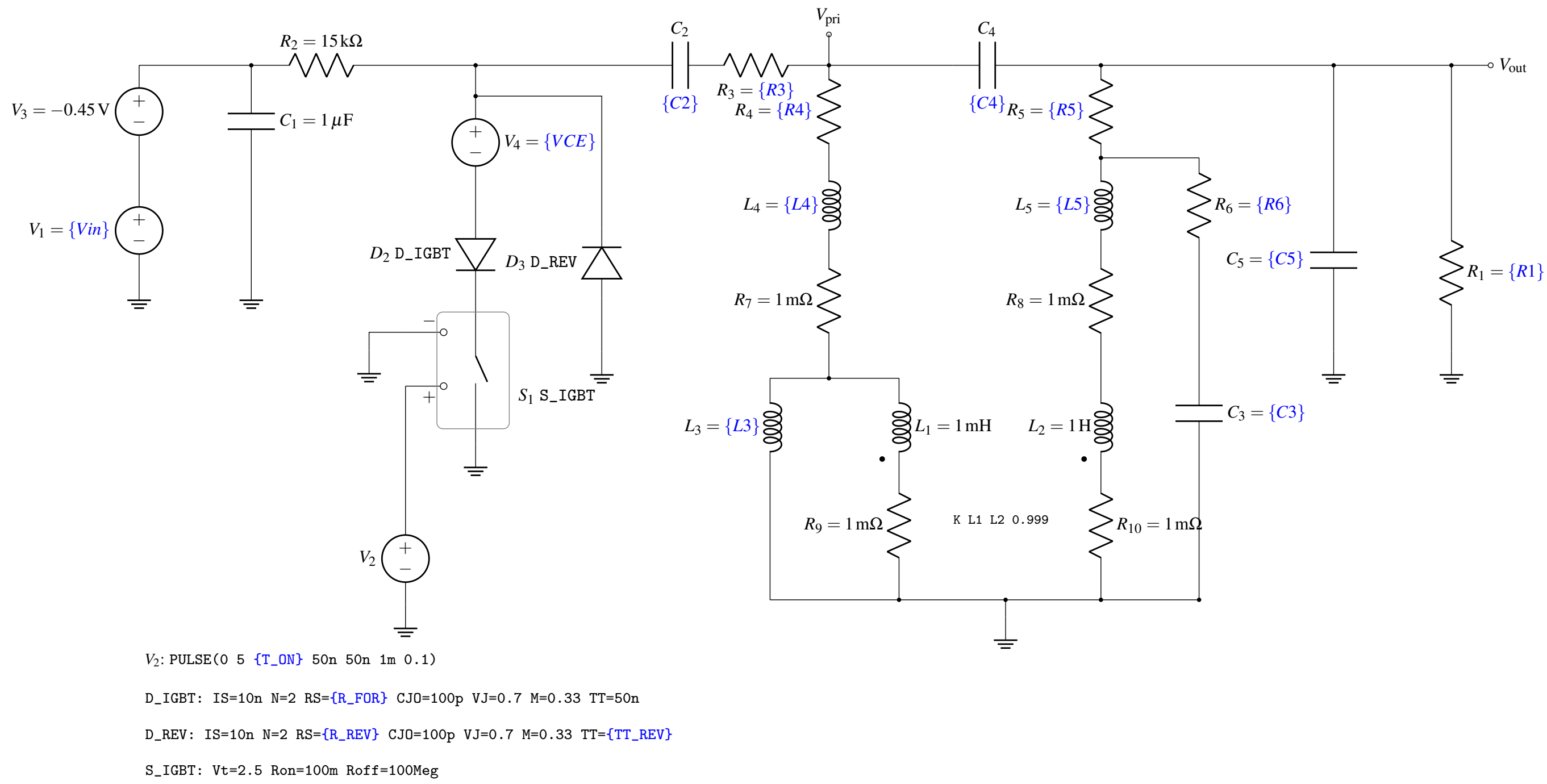

\subsection{Data Acquisition}
Waveforms for fitting the SPICE model were obtained in atmosphere, probed using a $100:1$ ratio high-voltage probe with a $1~\si{\kilo\volt}$ rating.
The oscilloscope probe had an input capacitance of $3.5~\si{\pico\farad}$, approximately the same as the capacitance of the DBD electrode measured at $100~\si{\kilo\hertz}$ using an LCR meter.
The provided traces were measured without the DBD electrode attached.
For the purposes of waveform fitting, it was assumed that the capacitive load of the DBD electrode and that of the oscilloscope probe were the same.
There is a slight discrepancy in the fitted capacitance (C5) and the measured load capacitance.
The TTL trigger pulse was supplied with a $10~\si{\hertz}$ repetition rate and also used as the trigger for the oscilloscope.
C2 was varied by soldering combinations of parallel capacitors into place using its PCB footprint.
A total of 7 different capacitances were tested, $47~\si{\nano\farad}$, $100~\si{\nano\farad}$, $147~\si{\nano\farad}$, $200~\si{\nano\farad}$, $300~\si{\nano\farad}$, $347~\si{\nano\farad}$, and $447~\si{\nano\farad}$.
However, the waveforms from the $100~\si{\nano\farad}$ and $347~\si{\nano\farad}$, measurements were randomly selected and excluded from the initial fit to use as unseen validation waveforms, leaving five waveforms that actually contributed to the fitted values.
A supply voltage of $15~\si{volt}$ was used to measure all of these reference traces.

\subsection{Equivalent Circuit Model Fit}
See fig. \ref{fig:equiv} to see the equivalent circuit that was fitted to obtain effective component values and extrapolations for the performance of the DBD drive.

The power supply is modeled with a $-0.45~\si{volt}$ offset representing the voltage drop from the 1N4007 diode at the input (intended to protect the circuit from reverse supply voltage).

The IGBT (ISL9V50363ST) is modelled as an idea voltage-controlled switch ($S_1$) with forward conduction modeled using $D_2$ with an additional degree of freedom for the collector-emitter voltage drop ($V_4$) and the inductive kickback protection diode (STTH810DI) diode is modeled with $D_3$.
Relatively generic parameters are used for both diode models because they were found to have little measurable effect on the simulation output.
The inclusion of $V_{CE}$ as a fitted parameter is intended as a catch-all correction for the non-linear diode characteristics that are not covered by the other two fit parameters, the series resistances.

For each of the recorded waveforms, $C_2$ and $R_3$ were fitted with the rated value of the capacitors used provided as a guess.
$R_3$ fitting allowed the varying ESR to be accounted for.

The transformer model is the most complex part of the equivalent circuit.
An ideal linear transformer model equivalent circuit model is used, built around an ideal transformer formed by $L_1$ and $L_2$.
Additional inductance representing leakage inductance ($L_4$ primary, $L_5$ secondary) and magnetizing inductance ($L_3$) are also included.
Resistance losses in the transformer are modelled by $R_4$ and $R_5$.
Interwinding capacitance is represented by $C_4$, and the combination of $C_3$ and $R_6$ represent the parasitic capacitance of the secondary.
The $1~\si{\milli\ohm}$ resistances are intended mainly to stabilize the simulations.
Without these, certain combinations of component values would cause it to fail to converge.
Finally, the leakage resistance and the capacitance of the DBD electrode were modelled using $C_5$ and $R_1$, loading the secondary.

The equivalent circuit model was developed through an iterative process of manually tuning effective component values and performing fits of the five input waveforms.
LTSpice was used for the design and initial manual tuning of an equivalent circuit model (see fig. \ref{fig:equiv}).
However, the process of running the simulations for the parameter-fitting routine was impractical to perform using LTSpice (the software could only be run from the Python fitting code by using the PyLTSpice library, which requires the GUI to be opened for each evaluation), so the PySpice library was used instead, with Ngspice version 46 used as the headless simulation backend.
The code is provided for reproducibility, but the fitting procedure itself is cumbersome and features several workarounds for bugs in the PySpice library.

The CMA-ES (Covariance Matrix Adaptation Evolution Strategy) algorithm, implemented in the Python library cma, was used to fit the component values.
The cost function $C(\vec{p})$ (with fit parameters $p_j$) is defined as
\begin{equation*}
    C(p_j)=\frac{1}{5}\sum_{i=1}^{5}\int\left(V_{meas,\ i}(p_j;\, t)-V_{model,\ i}(p_j;\, t)\right)^2\, dt\ ,
\end{equation*}
where the index $i$ indentifies the five waveforms used as the fit input.
A fit window of $15~\si{\micro\second}$ was used for all of the fits, including $1~\si{\micro\second}$ from before the oscilloscope is triggered.

The choice of the algorithm proceeded mostly by trial and failure, testing options that are derivative-free and support bounds for the fit parameters.
The fit is very prone to converging to local minima, so a manual initial fit that qualitatively overlaps with the experimental measurement is critical.
In particular, the slower oscillation of the primary should be manually fitted carefully; the secondary's frequency and decay are fitted well by the optimization algorithm.
See Table \ref{tab:equiv-fit} for the fitted values.

An unsuccessful effort was made to use several equivalent circuit models of DBDs found in the literature to reproduce the timings of the sharp fluorescence spikes that were measured during the experiment.
The non-linear DBD load is therefore not included in the simulation (and the transformer's saturation is also neglected), so the actual output waveforms during DBD operation should not be assumed to be what the linear model predicts.

\begin{table*}[htbp!]
\centering
\caption{Fitted equivalent-circuit parameters from the simultaneous fit to five
nominal $15\,\mathrm{V}$ measurements. Shared parameters were common to all
measurements; $C_2$, $R_3$, and $T_{\mathrm{ON}}$ were fitted independently for
each capacitor setup. Values are rounded to four significant figures.}
\alttext{Table of fitted circuit parameters. The upper part lists the component values shared by all measurements; the lower part lists the values fitted separately for each primary capacitor. The fitted primary capacitances agree with the nominal values to within about 10 percent.}
\label{tab:equiv-fit}
\begingroup
\small
\renewcommand{\arraystretch}{1.12}
\setlength{\tabcolsep}{9pt}
\begin{tabular}{@{}llrl@{}}
\hline
\multicolumn{4}{@{}l}{\textbf{Shared parameters}} \\
\hline
Parameter & Circuit role & Fitted value & Unit \\
\hline
$V_{\mathrm{in}}$ & Supply source $V_1$ & 14.99 & $\mathrm{V}$ \\
$V_{\mathrm{CE}}$ & Additional switch-branch offset $V_4$ & 0.1018 & $\mathrm{V}$ \\
$R_1$ & Load leakage resistance & 341.1 & $\mathrm{k}\Omega$ \\
$R_4$ & Primary series resistance & 0.5641 & $\Omega$ \\
$R_5$ & Secondary series resistance & 18.35 & $\Omega$ \\
$R_6$ & Resistance in the $R_6$--$C_3$ branch & 258.1 & $\Omega$ \\
$C_3$ & Transformer shunt capacitance & 9.616 & $\mathrm{pF}$ \\
$C_4$ & Interwinding capacitance & 0.1269 & $\mathrm{pF}$ \\
$C_5$ & Output shunt capacitance & 2.443 & $\mathrm{pF}$ \\
$L_3$ & Primary shunt inductance & 22.32 & $\mu\mathrm{H}$ \\
$L_4$ & Primary series inductance & 4.624 & $\mu\mathrm{H}$ \\
$L_5$ & Secondary series inductance & 143.4 & $\mu\mathrm{H}$ \\
$R_{\mathrm{FOR}}$ & \texttt{D\_IGBT} model series resistance & 8.863 & $\mathrm{m}\Omega$ \\
$R_{\mathrm{REV}}$ & \texttt{D\_REV} model series resistance & 1.286 & $\mathrm{m}\Omega$ \\
$TT_{\mathrm{REV}}$ & \texttt{D\_REV} model transit time & 15.47 & $\mathrm{ns}$ \\
\hline
\end{tabular}

\vspace{1em}
\begin{tabular}{@{}rrrr@{}}
\hline
\multicolumn{4}{@{}l}{\textbf{Parameters fitted separately for each setup}} \\
\hline
Nominal capacitance & $C_2$ & $R_3$ & $T_{\mathrm{ON}}$ \\
($\mathrm{nF}$) & ($\mathrm{nF}$) & ($\Omega$) & ($\mu\mathrm{s}$) \\
\hline
47 & 42.02 & 3.314 & 0.9000 \\
147 & 139.1 & 0.9503 & 0.9687 \\
200 & 196.5 & 0.5651 & 0.9680 \\
300 & 330.0 & 0.1624 & 0.9881 \\
447 & 455.4 & 0.3825 & 0.9772 \\
\hline
\end{tabular}

\vspace{0.7em}
\begin{minipage}{0.94\textwidth}
\footnotesize
$V_{\mathrm{in}}$ is the fitted value of source $V_1$, before the fixed
$-0.45\,\mathrm{V}$ offset from $V_3$.
$V_{\mathrm{CE}}$ is the additional series offset from $V_4$, not the total
switch-branch on-state voltage.
$T_{\mathrm{ON}}$ is the switch-trigger delay; the pulse width remains fixed
at $1\,\mathrm{ms}$.
$TT_{\mathrm{REV}}$ is the SPICE diode transit-time parameter, not directly
the datasheet reverse-recovery time.
\end{minipage}
\endgroup
\end{table*}

\section{Fluorescence Baseline Subtraction}\label{sec:fluor}
To qualitatively remove the slow-varying baseline of the fluorescence signal to qualitatively examine the sharp fluorescence spikes (caused by DBD micro-discharges), we performed an asymmetric least-squares fit \cite{eilers2005}.
The baseline $z_i$ of a measured signal $y_i=(y_1,\dots,y_m)$ is obtained by minimizing the penalized least-squares functional
\begin{equation}
  S(z_i) = \sum_{i=1}^{m} w_i\,(y_i - z_i)^2 + \lambda \sum_{i=4}^{m} \left(\Delta^{3} z_i\right)^2 ,
\end{equation}
where the first term measures the fidelity of $z_i$ to the data, and the second term penalizes the third-order finite differences $\Delta^{3}z_i$ of the baseline and thereby enforces smoothness.
The parameter $\lambda$ sets the balance between the two terms; large values yield a stiffer baseline. Setting $\partial S/\partial z_i=0$ gives the linear system
\begin{equation}
  \left(W_{ij} + \lambda\,D_{ij}D_{ij}^{\mathsf{T}}\right)z_i = W_{ij}y_i,
\end{equation}
where $W_{ij}=\mathrm{diag}(w_i)$ and $D_{ij}$ is the third-order difference matrix, such that $D_{ij}^{\mathsf{T}}z_i=\Delta^{3}z_i$.
The asymmetry is introduced through the weights.
Initially all $w_i = 1$, so the first iteration returns a smooth fit that passes through the peak.
The weights are then updated iteratively according to
\begin{equation}
  w_i =
  \begin{cases}
    p & \text{if}\  y_i > z_i \\
    1 - p & \text{if}\ y_i < z_i
  \end{cases}
\end{equation}
with $0<p\ll 1$, so that points lying above the current baseline estimate (the sharp peaks) are strongly suppressed while points below it are retained.
After each update, the system is solved again, and the baseline relaxes towards the lower envelope of the data and no longer follows the peak.
Here, we used $\lambda = 10^{6}$, $p = 10^{-4}$, and 30 iterations.
These values were chosen qualitatively, to cleanly isolate the fluorescence spikes for visual clarity.
In addition, the baseline was constrained to non-negative values after every iteration.

\makeatletter
\let\auto@bib@innerbib\@empty
\makeatother